\documentclass[12pt]{article}
\usepackage{graphicx}
\usepackage{lscape}
\usepackage{citesort}
\usepackage{amssymb}
\usepackage{appendix}
\usepackage{multirow}

\begin{document}

\def\Tr{\mbox{Tr}}
\def\figt#1#2#3{
        \begin{figure}
        $\left. \right.$
        \vspace*{-2cm}
        \begin{center}
        \includegraphics[width=10cm]{#1}
        \end{center}
        \vspace*{-0.2cm}
        \caption{#3}
        \label{#2}
        \end{figure}
	}
	
\def\figb#1#2#3{
        \begin{figure}
        $\left. \right.$
        \vspace*{-1cm}
        \begin{center}
        \includegraphics[width=10cm]{#1}
        \end{center}
        \vspace*{-0.2cm}
        \caption{#3}
        \label{#2}
        \end{figure}
                }

\def\ds{\displaystyle}
\def\beq{\begin{equation}}
\def\eeq{\end{equation}}
\def\bea{\begin{eqnarray}}
\def\eea{\end{eqnarray}}
\def\beeq{\begin{eqnarray}}
\def\eeeq{\end{eqnarray}}
\def\ve{\vert}
\def\vel{\left|}
\def\ver{\right|}
\def\nnb{\nonumber}
\def\ga{\left(}
\def\dr{\right)}
\def\aga{\left\{}
\def\adr{\right\}}
\def\lla{\left<}
\def\rra{\right>}
\def\rar{\rightarrow}
\def\lrar{\leftrightarrow}  
\def\nnb{\nonumber}
\def\la{\langle}
\def\ra{\rangle}
\def\ba{\begin{array}}
\def\ea{\end{array}}
\def\tr{\mbox{Tr}}
\def\ssp{{\Sigma^{*+}}}
\def\sso{{\Sigma^{*0}}}
\def\ssm{{\Sigma^{*-}}}
\def\xis0{{\Xi^{*0}}}
\def\xism{{\Xi^{*-}}}
\def\qs{\la \bar s s \ra}
\def\qu{\la \bar u u \ra}
\def\qd{\la \bar d d \ra}
\def\qq{\la \bar q q \ra}
\def\GG{\langle g_s^2 G^2 \rangle}
\def\q{\gamma_5 \not\!q}
\def\x{\gamma_5 \not\!x}
\def\g5{\gamma_5}
\def\sb{S_Q^{cf}}
\def\sd{S_d^{be}}
\def\su{S_u^{ad}}
\def\sbp{{S}_Q^{'cf}}
\def\sdp{{S}_d^{'be}}
\def\sup{{S}_u^{'ad}}
\def\ssp{{S}_s^{'??}}

\def\sig{\sigma_{\mu \nu} \gamma_5 p^\mu q^\nu}
\def\fo{f_0(\frac{s_0}{M^2})}
\def\ffi{f_1(\frac{s_0}{M^2})}
\def\fii{f_2(\frac{s_0}{M^2})}
\def\O{{\cal O}}
\def\sl{{\Sigma^0 \Lambda}}
\def\es{\!\!\! &=& \!\!\!}
\def\ap{\!\!\! &\approx& \!\!\!}
\def\ar{&+& \!\!\!}
\def\ek{&-& \!\!\!}
\def\kek{\!\!\!&-& \!\!\!}
\def\cp{&\times& \!\!\!}
\def\se{\!\!\! &\simeq& \!\!\!}
\def\eqv{&\equiv& \!\!\!}
\def\kpm{&\pm& \!\!\!}
\def\kmp{&\mp& \!\!\!}
\def\mcdot{\!\cdot\!}
\def\erar{&\rightarrow&}


\def\simlt{\stackrel{<}{{}_\sim}}
\def\simgt{\stackrel{>}{{}_\sim}}


\title{
         {\Large
                 {\bf
Electromagnetic form factors of octet baryons in QCD
                 }
         }
      }

\author{\vspace{1cm}\\
{\small T. M. Aliev \thanks {e-mail:
taliev@metu.edu.tr}~\footnote{permanent address:Institute of
Physics,Baku,Azerbaijan}\,\,, K. Azizi \thanks {e-mail:
kazizi@dogus.edu.tr}\,\,, M. Savc{\i} \thanks
{e-mail: savci@metu.edu.tr}} \\
{\small Physics Department, Middle East Technical University,
06531 Ankara, Turkey }\\
{\small$^\ddag$ Physics Department,  Faculty of Arts and Sciences,
Do\u gu\c s University,} \\
{\small Ac{\i}badem-Kad{\i}k\"oy,  34722 Istanbul, Turkey}}

\date{}

\begin{titlepage}
\maketitle
\thispagestyle{empty}

\begin{abstract}

The electromagnetic form factors of octet baryons are estimated within light
cone QCD sum rules method, using the most general form of the interpolating
current for baryons. A comparison of our predictions on the magnetic dipole
and electric form factors with the results of other approaches is performed.

\end{abstract}

~~~PACS numbers: 11.55.Hx, 14.20.--c, 14.20.Mr

\end{titlepage}

\section{Introduction}

Electromagnetic form factors of nucleons provide information on their
internal structures, i.e., about the spatial distribution of charge and
magnetization of the nucleon. Nucleon electromagnetic form factors that 
are the functions of only four--momentum transfer squared $Q^2$ are
described by Dirac $F_1(Q^2)$ and Pauli $F_2(Q^2)$ form factors which are
related to the electric and magnetic dipole form factors $G_E(Q^2)$ and
$G_M(Q^2)$ as,
\bea
\label{egegm01}
G_E \es F_1(Q^2) - {Q^2 \over 4 m_B^2} F_2(Q^2)~, \nnb \\
G_M \es F_1(Q^2) + F_2(Q^2)~.   
\eea
Obviously, in the limit $Q^2 \to 0$ the form factors $G_E$ and $G_M$
correspond to the charge and magnetic moments of the nucleon, while $F_1$ and
$F_2$ describe the charge and anomalous magnetic moments of the nucleon.

The study of electromagnetic form factors of hadrons receives a lot of
attention during the past decade.
Recent experiments on the nucleon form factors using the polarized electron
beam and polarized protons, which are presented in detail in
\cite{Rgegm01}, allow more accurate measurements of the nucleon
form factors at higher values of the momentum transfer. In the polarization
measurements it is observed that the ratio $G_E^P(Q^2)/G_M^P(Q^2)$ can not
be determined by the simple dipole form $G_D(Q^2) = (1+Q^2/Q_0^2)^{-2}$ with
$Q_0^2 = (0.71~GeV)^2$ \cite{Rgegm02,Rgegm03,Rgegm04}. The neutron form factors
that are measured up to $Q^2=3.4~GeV^2$ recently can provide detailed
comparison of the proton and neutron form factors \cite{Rgegm05,Rgegm06}.

Considerable progress has also been achieved on the electromagnetic excitation of
nucleon resonances during last years. The cross sections and photon asymmetries for the 
photo production of the pion and $\eta$ mesons are measured at MAMI at Mainz, ELSA
at Bonn, LEGS at Brookhaven, and GRAAL at Grenoble. Moreover, a large amount of
data has already been collected for the $\Delta(1232)$ excitation and
single $Q^2$ data points are obtained for the longitudinal and transversal
form factors of the $p \to \Delta (1232),P_{11}(1440),S_{11}(1535)$,
$D_{13}(1520)$, etc., whose results are all given in
\cite{Rgegm07}.
These progresses in experiments open a way to real possibility of measuring
the electromagnetic form factors of the octet baryons in near future.   

In the present work we calculate the electromagnetic form factors of the octet
baryons within the light cone QCD sum rules (LCSR) method by employing the general
form of the interpolating currents. It should be noted that this problem is
studied in the same method for the Ioffe current alone in
\cite{Rgegm08,Rgegm09,Rgegm10}. It should also be reminded to the interested
reader that the nucleon electromagnetic form factors are calculated for the
Ioffe and general currents in \cite{Rgegm11} and \cite{Rgegm12},
respectively.

The plan of this work is as follows. In section 2 we introduce the
correlation function which we shall use in calculating the electromagnetic
form factors of the octet baryons, and discuss how the interpolating currents
of the octet baryons are related to each other. In section 3, the light cone QCD
sum rules for the electromagnetic form factors are obtained in the case when
the correlation functions are calculated in terms of the main nonperturbative input
parameters, namely in terms of distribution amplitudes (DAs) of the octet
baryons. The last section
contains the details of the numerical calculations of the electromagnetic
form factors of the octet baryons.

\section{LCSR for the electromagnetic form factors of octet baryons}

In order to obtain the LCSR for the electromagnetic form factors of the
octet baryons we start by considering the following vacuum--to--one--octet
baryon correlation function,
\bea
\label{egegm02}
\Pi_\mu(p,q) = i \int d^4x e^{iqx} \lla 0 \vel T \Big\{\eta(0)
j_\mu^{el}(x) \Big\} \ver B(p) \rra~,
\eea
where $\eta$ is the interpolating current of the octet baryon, $j_\lambda^{el}(x)$
is the electromagnetic current, $\mu$ is the vector Lorentz index, $T$
is the time ordering operation, and $B(p)$ is the one particle baryon state
with momentum $p$.

The most general forms of the interpolating currents for the octet baryons can
be written as,
\bea
\label{egegm03}
\eta^{\Sigma^0} \es \sqrt{2} \varepsilon^{abc} \sum_{i=1}^2 \Big[
\Big( u^{aT} C A_1^\ell s^b \Big) A_2^\ell d^c +
\Big( d^{aT} C A_1^\ell s^b \Big) A_2^\ell u^c \Big]~, \nnb \\
\eta^{\Sigma^+} \es 2 \varepsilon^{abc} \Big( u^{aT} C A_1^\ell s^b
\Big)A_2^\ell u^c~, \nnb \\
\eta^{\Sigma^-} \es 2 \varepsilon^{abc} \Big( d^{aT} C A_1^\ell s^b
\Big)A_2^\ell d^c~, \nnb \\
\eta^{\Xi^0}    \es \eta^{\Sigma^+} (u \lrar s)~, \nnb \\
\eta^{\Xi^-}    \es \eta^{\Sigma^-} (d \lrar s)~,
\eea
where $A_1^1 = I$, $A_1^2 = A_2^1 = \gamma_5$, $A_2^2 = \beta$.

The interpolating current of the $\Lambda$ baryon can also be obtained from that
of $\Sigma^0$ baryon in the following way \cite{Rgegm13}:
\bea
\label{egegm04}
2 \eta^{\Sigma^0}(d \lrar s) + \eta^{\Sigma^0} = - \sqrt{3} \eta_\Lambda~,~\mbox{or,} \nnb \\
2 \eta^{\Sigma^0}(u \lrar s) - \eta^{\Sigma^0} = - \sqrt{3} \eta_\Lambda~.
\eea

Our primary aim  is the calculation of the phenomenological
part of the correlation function
(\ref{egegm02}). According to the standard procedure, in order to obtain the
physical part of the correlation function of 
the octet baryons we insert a full set of baryons into Eq.
(\ref{egegm02}). Separating the contribution of the ground state baryon we
get,
\bea
\label{egegm05}
\Pi_\mu (p,q) \es {\lla 0 \vel \eta \ver B(p-q) \rra \lla B(p-q) \vel
j_\mu^{el} \ver B(p) \rra \over m_{B}^2 - (p-q)^2 } + \cdots~,
\eea
where $\cdots$ represents the contributions of the higher states and
continuum.

The matrix element $\lla 0 \vel \eta \ver B(p-q) \rra$ appearing in Eq. (\ref{egegm05}) are determined as,
\bea
\label{egegm06}
\lla 0 \vel \eta \ver B(p-q) \rra \es \lambda_B u(p-q)~,
\eea
where $\lambda_B$ is the residue of the members of the octet baryons.
The hadronic matrix element with the electromagnetic current is determined in
terms of three independent form factors $F_1$, $F_2$ and $F_3$ in
the following way,
\bea
\label{egegm07}
\lla B(p-q) \vel j_\mu^{el} \ver B(p) \rra \es \bar{u} (p-q) \Big[
F_1(q^2) \gamma_\mu - i {\sigma_{\mu\nu} q^\nu \over 2 m_B} F_2(q^2) +
{q_\mu \over 2 m_B} F_3(q^2) \Big] u(p)~.
\eea
From conservation of the electromagnetic current we get $F_3(q^2)=0$. Taking
Dirac equation into account, one can show that the general decomposition of
the correlation function (\ref{egegm02}) contains six independent amplitudes
in the presence of the electromagnetic current,
\bea
\label{egegm08}
\Pi_\mu(p,q) = \Big[ \Pi_1 p_\mu + \Pi_2 p_\mu \rlap/{q} + \Pi_3
\gamma_\mu + \Pi_4 \gamma_\mu \rlap/{q} + \Pi_5 q_\mu + \Pi_6 q_\mu
\rlap/{q} \Big] u (p)~.
\eea
Using the definitions given by Eqs. (\ref{egegm06}) and (\ref{egegm07}),
we get the following expression for the hadronic
part,
\bea
\label{egegm09}
\Pi_\mu(p,q) \es {\lambda_B \over m_B^2 - (p-q)^2} \Bigg\{ 2 F_1(q^2) p_\mu
+ {F_2(q^2) \over m_B} p_\mu \rlap/{q} + \Big[F_1(q^2) + F_2(q^2) \Big]
\gamma_\mu \rlap/{q} \nnb \\
\ar \Big[ -2 F_1(q^2) - F_2(q^2) \Big] q_\mu - {F_2(q^2) \over 2 m_B} q_\mu
\rlap/{q} \Bigg\}~.
\eea
Equating the coefficients of each Lorentz structure in Eqs. (\ref{egegm08})
and (\ref{egegm09}) we get the sum rules for the form factors.
In order to perform numerical
analysis we need expressions of the invariant functions
$\Pi_i,~(i=1,\cdots,6)$ from QCD side.

The calculation of the invariant functions $\Pi_i$ from QCD side is carried
out when
the external momenta $p-q$ and $q$ are taken in deep Eucledian space, i.e.,
$(p-q)^2 \ll 0$ and $q^2 \ll 0$, which is necessary to perform operator
product expansion (OPE) near the light cone $x^2 \sim 0$. The OPE result can be
obtained as the sum over octet baryon distribution amplitudes (DAs) of growing
twist, which are the main non--perturbative inputs of the LCSR.

As has already been noted, the DAs of $\Sigma$, $\Xi$ and $\Lambda$  
are investigated in \cite{Rgegm08,Rgegm09,Rgegm10}. The DAs of the octet
baryons with $J^P=1^+/2$ are defined from the matrix element of the three--quark
operator between the vacuum and the baryon state $\ve B(p) \ra$, whose form
is given as,
\bea
\label{egegm10}
\varepsilon^{abc} \la 0 \vel q_{1\alpha}^a(a_1x)  q_{2\beta}^b(a_2x)
q_{3\gamma}^c(a_3x) \ver B(p) \ra~,
\eea
where $\alpha,\beta,\gamma$ are the Dirac indices, $a,b,c$ are the color
indices, and $a_i$ are positive numbers satisfying $a_1+a_2+a_3=1$. Using
the Lorentz covariance, as well as spin and parity of the baryons under
consideration, the matrix element
(\ref{egegm10}) can be decomposed as,
\bea
\label{egegm11}
4 \varepsilon^{abc} \la 0 \vel q_{1\alpha}^a(a_1x)  q_{2\beta}^b(a_2x) 
q_{3\gamma}^c(a_3x) \ver B(p) \ra = \sum_i {\cal F}_i
\Gamma_{1i}^{\alpha\beta} \Big(\Gamma_{2i} B(p)\Big)_\gamma~,
\eea
where $\Gamma_{1(2)i}$ are certain Dirac matrices, ${\cal F}_i={\cal S}_i,
{\cal P}_i,{\cal A}_i,{\cal V}_i$ and ${\cal T}_i$ are the DAs which do not
have definite twists. The DAs with definite twists are determined from,
\bea
\label{egegm12}
4 \varepsilon^{abc} \la 0 \vel q_{1\alpha}^a(a_1x)  q_{2\beta}^b(a_2x)
q_{3\gamma}^c(a_3x) \ver B(p) \ra = \sum_i F_i
\Gamma_{1i}^{\prime\alpha\beta} \Big(\Gamma_{2i}^\prime B(p)\Big)_\gamma~,
\eea
where $F_i=S_i,P_i,A_i,V_i,A_i$ and $T_i$. The Relations among these two sets of
DAs are given as,
\bea
\label{egegm13}
\begin{array}{ll}
{\cal S}_1 = S_1~,& (2 P \mcdot x) \, {\cal S}_2 = S_1 - S_2~, \nnb \\
{\cal P}_1 = P_1~,& (2 P \mcdot x) \, {\cal P}_2 = P_2 - P_1~, \nnb \\
{\cal V}_1 = V_1~,& (2 P \mcdot x) \, {\cal V}_2 = V_1 - V_2 - V_3~, \nnb \\
2{\cal V}_3 = V_3~,& (4 P \mcdot x) \, {\cal V}_4 =
- 2 V_1 + V_3 + V_4 + 2 V_5~,\nnb \\
(4 P\mcdot x) \, {\cal V}_5 = V_4 - V_3~,&
(2 P \mcdot x)^2 \, {\cal V}_6 = - V_1 + V_2 + V_3 + V_4 + V_5 - V_6~, \nnb \\
{\cal A}_1 = A_1~,& (2 P \mcdot x) \, {\cal A}_2 = - A_1 + A_2 - A_3~, \nnb \\
2 {\cal A}_3 = A_3~,&
(4 P \mcdot x) \, {\cal A}_4 = - 2 A_1 - A_3 - A_4 + 2 A_5~, \nnb \\
(4 P \mcdot x) \, {\cal A}_5 = A_3 - A_4~,&
(2 P \mcdot x)^2 \, {\cal A}_6 = A_1 - A_2 + A_3 + A_4 - A_5 + A_6~, \nnb \\
{\cal T}_1 = T_1~, & (2 P \mcdot x) \, {\cal T}_2 = T_1 + T_2 - 2T_3~, \nnb \\
2 {\cal T}_3 = T_7~,& (2 P \mcdot x) \, {\cal T}_4 = T_1 - T_2 - 2 T_7~, \nnb \\
(2 P\mcdot x) \, {\cal T}_5 = - T_1 + T_5 + 2 T_8~,&
(2 P \mcdot x)^2 \, {\cal T}_6 = 2 T_2 - 2 T_3 - 2 T_4 + 2 T_5 + 2 T_7 + 2 T_8~, \nnb \\
(4P \mcdot x) \, {\cal T}_7 = T_7 - T_8~, &
(2 P\mcdot x)^2 \, {\cal T}_8 = - T_1 + T_2 + T_5 - T_6 + 2 T_7 + 2 T_8~. \nnb
\end{array}
\eea
The complete decomposition of the DAs in Eq. (\ref{egegm11}) in terms of
${\cal S}_i,{\cal P}_i,{\cal A}_i,{\cal V}_i$ and ${\cal T}_i$ functions,
as well as the explicit expressions of DAs, can all be found in
\cite{Rgegm08,Rgegm09,Rgegm10,Rgegm11}.

Omitting the details of calculations of the theoretical part and equating
the coefficients of the Lorentz structures $p_\mu$, $p_\mu \rlap/{q}$ from
hadronic and theoretical parts, and performing the Borel transformation and
continuum subtraction over
the variable $(p-q)^2$, we get the following sum rules for the form factors,

\bea
\label{egegm14}
F_1(Q^2) \es {L \over 2 \lambda_B} \Bigg\{ \int_{x_0}^1 dx \Bigg(
- {\rho_2(x)\over x} + {\rho_4(x)\over x^2 M^2} - {\rho_6(x)\over 2 x^3 M^4} \Bigg)
e^{-{\bar{x} Q^2 \over x M^2} + {x m_B^2 \over M^2} } \nnb \\
\ar \Bigg[ {\rho_4 (x_0) \over Q^2 + x_0^2 m_B^2} - {1\over 2 x_0}
{\rho_6(x_0) \over (Q^2 + x_0^2 m_B^2) M^2} \nnb \\
\ar {1\over 2} {x_0^2 \over (Q^2 + x_0^2 m_B^2)} \Bigg(
{d\over dx_0} {\rho_6(x_0) \over x_0 (Q^2 + x_0^2 m_B^2) M^2}
\Bigg) \Bigg]e^{-(s_0-m_B^2)/ M^2}
\Bigg\} \\ \nnb \\ \nnb \\
\label{eslff13}
F_2(Q^2) \es 2 m_B F_1(Q^2) \Big(\rho_2(x) \to \rho_2^{'}(x),~
\rho_4(x) \to \rho_4^{'}(x),~\rho_6(x) \to \rho_6^{'}(x) \Big)~, 
\eea
where $M$ is the Borel parameter, $s={\ds {\bar{x}\over x}} Q^2 + (1-x)
m_B^2$,  $x_0$ is the solution of the equality $s=s_0$, $m_B$ is the mass of the
members of the octet baryons and $\bar{x}=1-x$. The factor $L$ in Eq.
(\ref{egegm14}) is the normalization factor  whose value for the members
of the octet baryons is determined as,
\bea
\label{nolabel}
L = \left\{ \begin{array}{cl}
{\ds 1\over \ds 2}          & \mbox{ for $\Sigma^+,\Sigma^-,\Xi^0~,\Xi^-$,} \\ \\
{\ds \sqrt{2}\over \ds 4}   & \mbox{ for $\Sigma^0$,} \\ \\
{\ds 1\over \ds 2 \sqrt{6}} & \mbox{ for $\Lambda$~.}
\end{array} \right.
\eea
The explicit expressions of
$\rho_i$ and $\rho_i^{'}$ for $\Sigma^+$, $\Sigma^0$ and $\Lambda$ baryons are
presented in the Appendix.

\section{Numerical analysis}

As has already been mentioned, 
the main nonperturbative inputs of LCSR are the baryon DAs. Here we would like to make the following remark about the expressions of the DAs
for the $\Lambda$, $\Sigma$ and $\Xi$ baryons. In \cite{bir}, the DAs
for nucleons were extended up to next-to leading order in conformal
spin and the expressions of the nucleon DAs of twist-3 up to next-to-next to
leading conformal spin is found in \cite{iki}. As a result of these two works it is
obtained that the nucleon form factors are sensitive to the higher conformal spin contributions. 
For other members of the octet baryons similar calculations are
not yet done and deserves a detailed study. In present work,  we consider the
DAs for the  $\Lambda$, $\Sigma$ and $\Xi$ baryons without these contributions, whose
expressions can be found in \cite{Rgegm08,Rgegm09,Rgegm10,Rgegm11}.
The parameters appearing in the
expressions of the DAs are estimated from the analysis of the sum rules
given in \cite{Rgegm10,Rgegm11,Rgegm12}:

\bea
\label{nolabel}
f_\Xi \es (9.9 \pm 0.4)\times 10^{-3}~GeV^2~, \nnb \\
\lambda_1 \es -(2.1 \pm 0.1)\times 10^{-2}~GeV^2~, \nnb \\ 
\lambda_2 \es (5.2 \pm 0.2)\times 10^{-2}~GeV^2~, \nnb \\
\lambda_3 \es (1.7 \pm 0.1)\times 10^{-2}~GeV^2~, \nnb \\ \nnb \\
f_\Sigma \es (9.4 \pm 0.4)\times 10^{-3}~GeV^2~, \nnb \\ 
\lambda_1 \es -(2.5 \pm 0.1)\times 10^{-2}~GeV^2~, \nnb \\
\lambda_2 \es (4.4 \pm 0.1)\times 10^{-2}~GeV^2~, \nnb \\  
\lambda_3 \es (2.0 \pm 0.1)\times 10^{-2}~GeV^2~, \nnb \\ \nnb \\
f_\Lambda \es (6.0 \pm 0.3)\times 10^{-3}~GeV^2~, \nnb \\ 
\lambda_1 \es (1.0 \pm 0.3)\times 10^{-2}~GeV^2~, \nnb \\
\vel \lambda_2 \ver \es (0.83 \pm 0.05)\times 10^{-2}~GeV^2~, \nnb \\  
\vel \lambda_3 \ver \es (0.83 \pm 0.05)\times 10^{-2}~GeV^2~. \nnb
\eea

The remaining input parameters of the LCSR are the continuum threshold
$s_0$, the Borel parameter $M^2$ and the auxiliary parameter $\beta$ that
appears in the expressions of the interpolating currents of the
octet baryons.

In our numerical calculations we use the values
$s_0=2.5~GeV^2$, $s_0=3.0~GeV^2$ and  $s_0=3.2~GeV^2$  for the continuum
threshold, obtained from mass 
sum rules analysis in \cite{Rgegm14}, for the $\Lambda$, $\Sigma$ and $\Xi$ baryons,
respectively.

The Borel mass parameter $M^2$ is another auxiliary parameter of the sum rules.
Therefore the ``working region" of $M^2$ should be found, where the form
factors are practically independent of it. The lower limit of $M^2$ can be
obtained by requiring that the higher states and continuum contributions to
the sum rules constitute, maximally, about 40\% of the total result. The
upper limit of $M^2$ can be determined by demanding that the operator product
expansion should be convergent. Our calculations show that the region in
which the aforementioned two conditions are properly satisfied, are:
$1.2~GeV^2 \le M^2 \le 3.0~GeV^2$ for $\Sigma$ and $\Lambda$ baryons; and
$1.4~GeV^2 \le M^2 \le 3.5~GeV^2$ for $\Xi$ baryons. In
further numerical analysis, we use $M^2=2.0~GeV^2$ for $\Sigma$, $\Xi$ and
$\Lambda$ baryons.   

The residues of the octet baryons are calculated in \cite{Rgegm14} and we
shall use these results in our numerical analysis. Furthermore, it should be
noted that, from experimental point of view, it is more convenient to study
the Sachs form factors $G_E$ and $G_M$ that are given in Eq.
(\ref{egegm01}).

The $Q^2$ dependence of the magnetic and electric form factors for
$\Sigma^+$, $\Xi^-$ and $\Lambda$,  baryons are shown in Figs. (1)--(6). In order
to get ``good" convergence of the light cone expansion and reliable results
from the LCSR, sufficiently large $Q^2$ is needed.
In our numerical calculations we consider the lower limit of $Q^2$ as
$Q^2=1~GeV^2$, where above this point the higher twist
contributions are suppressed. On the other side, the higher resonance and
continuum contributions become small enough when $Q^2 \le 8~GeV^2$. For this
reason, we perform numerical analysis in the region $1~GeV^2 \le Q^2 \le 8~GeV^2$.

In odd--numbered Figs.: (1), (3), (5) (even--numbered Figs.: (2), (4),
(6)) we present the dependence of the magnetic
dipole form factors $G_M(Q^2)$ (electric form factor $G_E(Q^2)$) on $Q^2$,
at fixed values of $s_0$ and $M^2$ chosen from their working regions,
and at several fixed values of the
arbitrary parameter $\beta$, for  $\Sigma^+$, $\Xi^-$ and $\Lambda$
baryons, respectively.

\begin{itemize}

\item In the case of $\Sigma^+$, these form factors get positive (negative)
values for negative (positive) values of the parameter $\beta$.

\item The situation for $\Sigma^-$ is contrary to the behaviors of the form
factors of the $\Sigma^+$, i.e., the values of $G_M$ and $G_E$ are positive
(negative) when the parameter $\beta$ is positive (negative).

\item In the case of $\Sigma^0$ baryon, the form factors exhibit similar
behaviors as the form factors of $\Sigma^+$ baryon.

\item It is observed that the form factor $G_E$ of the $\Xi^0$ baryon
changes its sign practically at all considered values of $\beta$.
The zero values of
$G_E$ depend on the values of the arbitrary parameter $\beta$. But the
values of $G_E$ are quite small, whose maximum value is about $G_E(Q^2) =
0.06$.

\item It is interesting to observe that at $\beta=3$ and $\beta=5$,
$G_M$ for the $\Xi^0$ changes its sign, while for the other values of
$\beta$ it is always negative.

\item For positive (negative) values of $\beta$ the magnetic dipole form
factor $G_M(Q^2)$ for $\Lambda$ baryon attains at positive (negative)
values.

\item The situation for electric form factor $G_E(Q^2)$, however, is
slightly different. Namely, in the case of Ioffe current for which
$\beta=-1$, $G_E(Q^2)$ becomes negative whose magnitude is negligibly small.

\item When the form factors of the $\Xi^-$ baryon are considered we see that
$G_M$ changes its sign only at $\beta=5$, while for all other values of
$\beta$ it gets at only negative values. On the other hand the form factor $G_E$
is positive (negative) for all positive and negative values of $\beta$.
 
\end{itemize}

We now compare our results on $Q^2$ dependence of the magnetic and charge
form factors with the ones existing in the literature. These form factors are
discussed within the LCSR method for the Ioffe current ($\beta = -1$)    
in \cite{Rgegm09,Rgegm10}, within the framework of the fully relativistic
constituent quark model in \cite{Rgegm15}, in the framework of the
covariant spectator quark model \cite{Rgegm16} and lattice QCD
\cite{Rgegm17}.
When we compare our predictions on the form factors with the results of
the above--mentioned works we obtain that, at $\beta=-1$, our predictions on
$G_E$ are very close to the results of \cite{Rgegm09,Rgegm10}, and
\cite{Rgegm15}, except for the $\Sigma^0$ baryon.
Our predictions for the magnetic form factor $G_M$ agree within the errors
with the existing results. The small differences among the predictions can be
attributed to the different values of the input parameters used in the
numerical analysis.

As has already been noted, the interpolating currents of the octet baryons
contain also the auxiliary arbitrary parameter $\beta$. Obviously, the physically
measurable quantities should be independent of this parameter. In order to
find the working region of the parameter $\beta$ we demand that the form
factors are practically independent of it. As an example, in Figs.
(7) and (8) we present the dependence of $G_M(Q^2)$ and $G_E(Q^2)$ on
$\cos\theta~(\beta = \tan\theta)$
for $\Sigma^+$ baryon at fixed values of $s_0=3.0~GeV^2$ and
$M^2=2.0~GeV^2$, for two fixed choices of $Q^2$, namely, $Q^2=2.0~GeV^2$ and
$Q^2=3.0~GeV^2$. We see from these figures that, in
the region $-0.3 \le \cos\theta \le 0.3$, $G_E$ and $G_M$ show very weak
dependence on $\beta$. In other words, the working region of the parameter
$\beta$ for the $\Sigma^+$ baryon is $-0.3 \le \cos\theta \le 0.3$.

We perform similar analysis for all other members of the octet baryons and
find out that the region $-0.2 \le \cos\theta \le 0.2$ is the common working
region to them as well. It should be noted here that $G_M(Q^2)$ for
$\Xi^0$ and $\Xi^-$ baryons exhibit stability in the range $-0.2 \le \cos\theta
\le 0.2$.  Also note that $\beta=-1$ point, which is the Ioffe
current corresponding to $\cos\theta=-0.71$, belongs to the region where the
predictions for the form factors are not reliable. Choosing the values of
$M^2$ and $\beta$ from the relevant working regions, and from a comparison
of our predictions on the form factors with the results of the
above--mentioned works we see that,
\begin{itemize}

\item Predictions of all works for $G_M(Q^2)$ are very close to each other within
the error limits.

\item Our predictions on $G_E(Q^2)$ agree with the results of other approaches,
except for the $\Sigma^0$ and $\Lambda$ baryons. In these cases our results
are very close to the predictions of the lattice QCD, while
considerable disagreements are observed with those obtained in
\cite{Rgegm15} and \cite{Rgegm16}. 

\end{itemize}

The results obtained in this work can be
improved by taking into account the ${\cal O}(\alpha_s)$ corrections to the
distribution amplitudes, and more accurate values of the input parameters
entering the sum rules.

In conclusion, in the present work we have studied the charge $G_E(Q^2)$ and
magnetic dipole $G_M(Q^2)$ form factors within the LCSR method by using the
most general for of the interpolating currents for the octet baryons.
We have compared our predictions on these form factors
with the results existing in literature that were obtained in framework of
the relativistic quark model, lattice QCD and LCSR for the Ioffe
current.

\newpage

\appendix  

\section*{Appendix}  
\setcounter{equation}{0}

In thhis Appendix we present the explicit expressions of the functions
$\rho_2(\rho_2^{'})$, $\rho_4(\rho_4^{'})$ and $\rho_6(\rho_6^{'})$ entering
to the sum rules for the form factors $F_1(Q^2)$ and $F_2(Q^2)$, 

\bea
\label{Sigma+}
\rho_6^{\Sigma^+}(x) \es 4 e_u m_{\Sigma^+}^3 (1+\beta) x (m_{\Sigma^+}^2 x^2 + Q^2)
\;\check{\!\check{B}}_6 (x) \nnb \\
\ar 4 e_s m_{\Sigma^+}^2 \Big\{ m_{\Sigma^+}^2 m_s (1-\beta) x^2 \; \widehat{\!\widehat{C}}_6
+ (1+\beta) \Big[ m_{\Sigma^+} (m_{\Sigma^+}^2 x^2 + Q^2) x
\; \widehat{\!\widehat{B}}_6 \nnb \\
\ek m_s ( Q^2 \; \widehat{\!\widehat{B}}_6 + 2 m_{\Sigma^+}^2 x^2
\;\widehat{\!\widehat{B}}_8) \Big]\Big\} (x)~, \nnb \\ \nnb \\
\rho_4^{\Sigma^+} (x)\es
e_u m_{\Sigma^+}\big\{
- 2 m_{\Sigma^+}^2 x \Big[ (1-\beta) (
\check{\!\check{C}}_6+\check{\!\check{D}}_6)
 - (1+\beta) (2 \; \check{\!\check{B}}_6 - 3 \;\check{\!\check{B}}_8) \Big] (x)
\nnb \\
\ar (1-\beta) \Big[ m_{\Sigma^+}^2 x^2
(\check{D}_4 - 3 \check{D}_5 -\check{C}_4 + 3 \check{C}_5 )
+ 2 Q^2 (\check{D}_2 + \check{C}_2) \Big] (x) \nnb \\
\ar (1+\beta) \Big[ Q^2 (\check{B}_2 + 5                   
\check{B}_4)  - m_{\Sigma^+}^2 x^2 ( 2 \check{H}_1 
- 2 \check{E}_1 + \check{B}_2 -
\check{B}_4 
+ 6 \check{B}_5 + 12 \check{B}_7 ) \Big] (x) \nnb \\
\ek 2 m_{\Sigma^+}^2 x \int_0^{\bar{x}}dx_3\,\Big[ (1-\beta) ({A}_1^M -
{V}_1^M) + 3
(1+\beta) T_1^M\Big] (x,1-x-x_3,x_3) \Big\}\nnb \\
\ar e_d m_{\Sigma^+} \Big\{
- 2 m_{\Sigma^+}^2 x \Big[ (1-\beta) (\widetilde{\!\widetilde{C}}_6-
\widetilde{\!\widetilde{D}}_6)
 + 2 (1+\beta) \; \widetilde{\!\widetilde{B}}_8 \Big] (x) \nnb \\
\ar (1-\beta) \Big[ - m_{\Sigma^+}^2 x^2 (\widetilde{D}_4
- \widetilde{D}_5 + \widetilde{C}_4 - \widetilde{C}_5 ) \Big] (x) \nnb \\
\ar (1+\beta) \Big[ 2 Q^2 (\widetilde{B}_2 +
\widetilde{B}_4)  - 4 m_{\Sigma^+}^2 x^2 ( \widetilde{B}_5
+ 2 \widetilde{B}_7 ) \Big] (x) \nnb \\
\ek  2 m_{\Sigma^+}^2 x \int_0^{\bar{x}} \,dx_1 \Big[ (1-\beta)
({A}_1^M+{V}_1^M) + 2 (1+\beta)
T_1^M\Big] (x_1,x,1-x_1-x) \Big\} \nnb \\
\ar e_s m_{\Sigma^+} \Big\{
2 m_{\Sigma^+} (1+\beta) \Big[ m_{\Sigma^+} x
(2  \; \widehat{\!\widehat{B}}_6 - \widehat{\!\widehat{B}}_8 ) - m_s \;
\widehat{\!\widehat{B}}_6 \Big] (x) \nnb \\
\ar (1-\beta) \Big[
2 (m_{\Sigma^+}^2 x^2 \widehat{C}_5 + Q^2 \widehat{C}_2) -
m_{\Sigma^+} m_s x ( 2 \widehat{C}_2 - \widehat{C}_4 -
\widehat{C}_5 ) \Big] (x) \nnb \\
\ek (1+\beta) \Big[ Q^2 ( \widehat{B}_2 - 3 \widehat{B}_4) +
m_{\Sigma^+}^2 x^2 ( \widehat{B}_2 - \widehat{B}_4 + 2 \widehat{B}_5 +
4 \widehat{B}_7)
- 4 m_{\Sigma^+} m_s x ( \widehat{B}_4 -
\widehat{B}_5 ) \Big] (x) \nnb \\
\ek 2 m_{\Sigma^+}^2 (1+\beta) x \int_0^{\bar{x}}dx_1\, T_1^M (x_1,1-x_1-x,x)
\Big\}~, \nnb \\ \nnb \\
\rho_2^{\Sigma^+}(x) \es
2 e_u  m_{\Sigma^+} x \int_0^{\bar{x}}dx_3\,\Big[ (1-\beta) (A_1+ 2 A_3 - {V}_1
+2 {V}_3) \nnb \\
\ek (1+\beta)(P_1 + S_1 + 3 T_1 - 6 T_3) \Big] (x,1-x-x_3,x_3) \nnb \\
\ar 2 e_d m_{\Sigma^+} \Big\{
\Big[ (1-\beta) ( \widetilde{D}_2 - \widetilde{C}_2) +
(1+\beta) ( \widetilde{B}_2 - \widetilde{B}_4) \Big] (x) \nnb \\
\ek x \int_0^{\bar{x}} \,dx_1 \Big[ (1-\beta) (A_1+A_3 + V_1-V_3)
+ 2 (1+\beta)(T_1 - 2 T_3) \Big] (x_1,x,1-x_1-x) \Big\} \nnb \\
\ar 2 e_s \Big\{
m_{\Sigma^+} \Big[ (1-\beta) \widehat{C}_2 +  
(1+\beta)  ( \widehat{B}_2 - \widehat{B}_4) \Big] (x) \nnb \\
\ar \int_0^{\bar{x}}dx_1\,\Big[ (1-\beta) ( m_{\Sigma^+} x {V}_3 + m_s {V}_1 ) \nnb \\
\ek (1+\beta) \Big( m_{\Sigma^+} x (P_1 + S_1 + T_1 - 2 T_3) + m_s T_1 
\Big)\Big] (x_1,1-x_1-x,x) \Big\}~, \nnb \\ \nnb \\
\rho_6^{'\Sigma^+} (x) \es - 4 e_u m_{\Sigma^+}^2 (1+\beta) (m_{\Sigma^+}^2 x^2 + Q^2)  
\;\check{\!\check{B}}_6 (x) \nnb \\
\ek 4 e_s m_{\Sigma^+}^2  \Big\{ m_{\Sigma^+} m_s (1-\beta) x \; \widehat{\!\widehat{C}}_6
+ (1+\beta) \Big[ (m_{\Sigma^+}^2 x^2 + Q^2) \; \widehat{\!\widehat{B}}_6 +
m_{\Sigma^+} m_s x (\; \widehat{\!\widehat{B}}_6 - 2 \;
\widehat{\!\widehat{B}}_8 )
\Big]\Big\} (x)~, \nnb \\ \nnb \\
\rho_4^{'\Sigma^+} (x) \es
e_u m_{\Sigma^+}^2 \Big\{
- 3 (1+\beta) \;\check{\!\check{B}}_6 (x) \nnb \\
\ar x \Big[ (1-\beta) (2 \check{D}_2 - \check{D}_4 + 3
\check{D}_5 + 2\check{C}_2 +
\check{C}_4 - 3 \check{C}_5 ) \nnb \\
\ar 2 (1+\beta) (\check{H}_1 - \check{E}_1 + \check{B}_2 + 2 \check{B}_4 +3
\check{B}_5 + 6 \check{B}_7 ) \Big] (x) \Big\} \nnb \\
\ar e_d m_{\Sigma^+}^2 \Big\{
2  (1+\beta) \;\widetilde{\!\widetilde{B}}_6 (x) \nnb \\
\ar x \Big[ (1-\beta) (\widetilde{D}_4 - \widetilde{D}_5 +
\widetilde{C}_4 -  \widetilde{C}_5 ) +
2 (1+\beta) (\widetilde{B}_2 + \widetilde{B}_4 
+ 2 \widetilde{B}_5 + 4 \widetilde{B}_7 ) \Big] (x) \nnb \\
\ar 2 (1-\beta) \int_0^{\bar{x}} \, (A_1^M + {V}_1^M) (x_1,x,1-x_1-x) \Big\}\nnb \\
\ar e_s  m_{\Sigma^+} \Big\{
- 5 m_{\Sigma^+} (1+\beta) \;\widehat{\!\widehat{B}}_6 (x) \nnb \\
\ek 2  \Big[ (1-\beta) \Big(m_{\Sigma^+} x \widehat{C}_5 - 
( m_{\Sigma^+} x + m_s) \widehat{C}_2 \Big) -
(1+\beta) \Big( m_{\Sigma^+} x (\widehat{B}_4 + \widehat{B}_5 + 2 \widehat{B}_7) -
m_s ( \widehat{B}_2 + \widehat{B}_4) \Big) \Big] (x) \nnb \\
\ar 2  m_{\Sigma^+} (1-\beta) \int_0^{\bar{x}}dx_1\, (A_1^M - {V}_1^M) (x_1,1-x_1-x,x)
\Big\}~,
\nnb \\ \nnb \\
\rho_2^{'\Sigma^+} (x) \es 2 e_d (1-\beta) \int_0^{\bar{x}} \,(A_1 + {V}_1) (x_1,x,1-x_1-x) \nnb \\
\ek 2 e_s (1-\beta) \int_0^{\bar{x}}dx_1\,{V}_1 (x,1-x-x_3,x_3)~,\nnb \\ \nnb \\
%
%
\label{sigma0}
\rho_6^{\Sigma^0}(x) \es
4 e_u m_{\Sigma^0}^3 (1+\beta) x (m_{\Sigma^0}^2 x^2 + Q^2)
\;\check{\!\check{B}}_6 (x)
+ 4 e_d m_{\Sigma^0}^3 (1+\beta) x (m_{\Sigma^0}^2 x^2 + Q^2)
\;\widetilde{\!\widetilde{B}}_6 (x) \nnb \\
\ar 8 e_s m_{\Sigma^0}^2 \Big\{ m_{\Sigma^0}^2 m_s (1-\beta) x^2\; \widehat{\!\widehat{C}}_6
+ (1+\beta) \Big[ m_{\Sigma^0} x (m_{\Sigma^0}^2 x^2 + Q^2)
\; \widehat{\!\widehat{B}}_6 \nnb \\
\ek m_s ( Q^2 \; \widehat{\!\widehat{B}}_6 + 2 m_{\Sigma^0}^2 x^2
\;\widehat{\!\widehat{B}}_8) \Big]\Big\} (x)~, \nnb \\ \nnb \\
\rho_4^{\Sigma^0} (x)\es
e_u m_{\Sigma^0} \Big\{
- 2 m_{\Sigma^0}^2 x \Big[ 2 (1-\beta) \; \check{\!\check{C}}_6
 - (1+\beta) (2 \; \check{\!\check{B}}_6 - 5 \;\check{\!\check{B}}_8) \Big] (x) \nnb \\
\ar \Big[ 2 (1-\beta) \Big( m_{\Sigma^0}^2 x^2
(\check{D}_5
- \check{C}_4 + 2 \check{C}_5 )
- Q^2 (\check{D}_2 - \check{C}_2) \Big) \nnb \\
\ar (1+\beta) \Big( Q^2 (3 \check{B}_2 + 7                   
\check{B}_4)  + m_{\Sigma^0}^2 x^2 ( 2 \check{H}_1 
- 2 \check{E}_1 - \check{B}_2 +                
\check{B}_4 
- 10 \check{B}_5 - 20 \check{B}_7 ) \Big) \Big] (x) \nnb \\
\ek 2 m_{\Sigma^0}^2 x \int_0^{\bar{x}}dx_3\,\Big[ 2 (1-\beta) {V}_1^M + 5 (1+\beta)
T_1^M\Big] (x,1-x-x_3,x_3) \Big\}\nnb \\
\ar e_d m_{\Sigma^0} \Big\{
- 2 m_{\Sigma^0}^2 x \Big[ 2 (1-\beta) \; \widetilde{\!\widetilde{C}}_6
 - (1+\beta) (2 \; \widetilde{\!\widetilde{B}}_6 - 
5 \; \widetilde{\!\widetilde{B}}_8) \Big] (x) \nnb \\
\ar \Big[ (1-\beta) \Big( - 2 m_{\Sigma^0}^2 x^2 (\widetilde{D}_5
+ \widetilde{C}_4 - 2 \widetilde{C}_5 )  
+ Q^2 (\widetilde{D}_2 + \widetilde{C}_2) \Big) \nnb \\
\ar (1+\beta) \Big( Q^2 (3 \widetilde{B}_2 + 7
\widetilde{B}_4)  - m_{\Sigma^0}^2 x^2 ( 2 \widetilde{H}_1
- 2 \widetilde{E}_1 + \widetilde{B}_2 -
\widetilde{B}_4
+  10 \widetilde{B}_5 + 20 \widetilde{B}_7 ) \Big) \Big] (x) \nnb \\
\ek  2 m_{\Sigma^0}^2 x \int_0^{\bar{x}} \,dx_1 \Big[ 2 (1-\beta) {V}_1^M + 5 (1+\beta)
T_1^M\Big] (x_1,x,1-x_1-x) \Big\}\nnb \\
\ar 2 e_s m_{\Sigma^0} \Big\{
2 m_{\Sigma^0} (1+\beta) \Big[ m_{\Sigma^0} x
(2  \; \widehat{\!\widehat{B}}_6 - \widehat{\!\widehat{B}}_8 ) - m_s \;
\widehat{\!\widehat{B}}_6 \Big] (x) \nnb \\
\ar \Big[ (1-\beta) \Big(
2 (m_{\Sigma^0}^2 x^2 \widehat{C}_5 + Q^2 \widehat{C}_2) -
m_{\Sigma^0} m_s x ( 2 \widehat{C}_2 - \widehat{C}_4 -
\widehat{C}_5 ) \Big) \nnb \\
\ek (1+\beta) \Big( Q^2 ( \widehat{B}_2 - 3 \widehat{B}_4) +
m_{\Sigma^0}^2 x^2 ( \widehat{B}_2 - \widehat{B}_4 + 2 \widehat{B}_5 +
4 \widehat{B}_7)
- 4 m_{\Sigma^0} m_s x ( \widehat{B}_4 -
\widehat{B}_5)\Big) \Big] (x) \nnb \\
\ek 2m_{\Sigma^0}^2 (1+\beta) x \int_0^{\bar{x}}dx_1\, T_1^M (x_1,1-x_1-x,x) \Big\}~, \nnb \\ \nnb \\
\rho_2^{\Sigma^0}(x) \es
 - 2 e_u m_{\Sigma^0} \Big\{
\Big[ (1-\beta) ( \check{D}_2 + \check{C}_2) -
(1+\beta) ( \check{B}_2 - \check{B}_4) \Big] (x) \nnb \\
\ar x \int_0^{\bar{x}}dx_3\,\Big[ (1-\beta) (A_3 + 2 {V}_1 - 3 {V}_3) -
(1+\beta)(P_1 + S_1 - 5 T_1 + 10 T_3) \Big] (x,1-x-x_3,x_3) \Big\} \nnb \\
\ar 2 e_d m_{\Sigma^0} \Big\{
\Big[ (1-\beta) ( \widetilde{D}_2 - \widetilde{C}_2) +
(1+\beta) ( \widetilde{B}_2 - \widetilde{B}_4) \Big] (x) \nnb \\
\ar x \int_0^{\bar{x}} \,dx_1 \Big[ (1-\beta) (A_3 - 2 {V}_1 + 3 {V}_3) -
(1+\beta)(P_1 + S_1 + 5 T_1 - 10 T_3) \Big] (x_1,x,1-x_1-x) \Big\} \nnb \\
\ar 4 e_s \Big\{
m_{\Sigma^0} \Big[ (1-\beta) \widehat{C}_2 -  
(1+\beta)  ( \widehat{B}_2 - \widehat{B}_4) \Big] (x) \nnb \\
\ar \int_0^{\bar{x}}dx_1\,\Big\{ (1-\beta) ( m_{\Sigma^0} x {V}_3 + m_s {V}_1 )
+ (1+\beta) \Big[ 2 m_{\Sigma^0} x T_3 - (m_{\Sigma^0} x + 2 m_s) T_1 \Big]
\Big\} (x_1,1-x_1-x,x)\Big\}~, \nnb \\ \nnb \\
\rho_6^{'\Sigma^0} (x) \es
 - 4 e_u m_{\Sigma^0}^2 (1+\beta) (m_{\Sigma^0}^2 x^2 + Q^2)  
\;\check{\!\check{B}}_6 (x)
- 4 e_d m_{\Sigma^0}^2 (1+\beta) (m_{\Sigma^0}^2 x^2 + Q^2)  
\;\widetilde{\!\widetilde{B}}_6 (x) \nnb \\
\ek 8 e_s m_{\Sigma^0}^2  \Big\{ m_{\Sigma^0} m_s (1-\beta) x \; \widehat{\!\widehat{C}}_6
+ (1+\beta) \Big[ (m_{\Sigma^0}^2 x^2 + Q^2) \; \widehat{\!\widehat{B}}_6 +
m_{\Sigma^0} m_s x (\; \widehat{\!\widehat{B}}_6 - 2 \; \widehat{\!\widehat{B}}_8 )
\Big]\Big\} (x)~, \nnb \\ \nnb \\
\rho_4^{'\Sigma^0} (x) \es
- e_u m_{\Sigma^0}^2 \Big\{
(1+\beta) \;\check{\!\check{B}}_6 (x) \nnb \\
\ar 2 x \Big[ (1-\beta) (\check{D}_2 + \check{D}_5 - \check{C}_2 -
\check{C}_4 + 2 \check{C}_5 ) +
(1+\beta) (\check{H}_1 - \check{E}_1 - 2 \check{B}_2 - 3 \check{B}_4 - 5
\check{B}_5 - 10 \check{B}_7 ) \Big] (x) \nnb \\
\ar 2 (1-\beta) \int_0^{\bar{x}}dx_3\, (A_1^M - {V}_1^M) (x,1-x-x_3,x_3)\Big\} \nnb \\
\ar e_d m_{\Sigma^0}^2 \Big\{ 
- (1+\beta) \;\widetilde{\!\widetilde{B}}_6 (x) \nnb \\
\ar 2 x \Big[ (1-\beta) (\widetilde{D}_2 + \widetilde{D}_5 + \widetilde{C}_2 + 
\widetilde{C}_4 - 2 \widetilde{C}_5 ) +
(1+\beta) (\widetilde{H}_1 - \widetilde{E}_1 + 2 \widetilde{B}_2 + 3 \widetilde{B}_4 
+ 5 \widetilde{B}_5 + 10 \widetilde{B}_7 ) \Big] (x) \nnb \\
\ar 2  (1-\beta) \int_0^{\bar{x}}dx_1 \, (A_1^M + {V}_1^M) (x_1,x,1-x_1-x)\Big\} \nnb \\
\ek 2 e_s  m_{\Sigma^0} \Big\{ 
5 m_{\Sigma^0} (1+\beta) \;\widehat{\!\widehat{B}}_6 (x) \nnb \\
\ar 2 \Big[ (1-\beta) \Big( m_{\Sigma^0} x \widehat{C}_5 - 
( m_{\Sigma^0} x + m_s) \widehat{C}_2 \Big) -
(1+\beta) \Big( m_{\Sigma^0} x (\widehat{B}_4 + \widehat{B}_5 + 2 \widehat{B}_7) -
m_s ( \widehat{B}_2 + \widehat{B}_4) \Big) \Big] (x) \nnb \\
\ar 2 m_{\Sigma^0} (1-\beta) \int_0^{\bar{x}}dx_1\, {V}_1^M (x_1,1-x_1-x,x)\Big\}~, \nnb \\ \nnb \\
\rho_2^{'\Sigma^0} (x) \es - 2 e_u (1-\beta) \int_0^{\bar{x}}dx_3\, (A_1 - {V}_1) (x,1-x-x_3,x_3) \nnb \\
\ar 2 e_d (1-\beta) \int_0^{\bar{x}}dx_1 \,(A_1 + {V}_1) (x_1,x,1-x_1-x) \nnb \\
\ek 4 e_s (1-\beta) \int_0^{\bar{x}}dx_1\,{V}_1 (x,1-x-x_3,x_3)~, \nnb \\ \nnb \\
%
%
\label{Lambda}
\rho_6^{\Lambda}(x) \es - 12 e_u m_\Lambda^3 (1+\beta) x (m_\Lambda^2 x^2 + Q^2)
\;\check{\!\check{B}}_6 (x)
-20 e_d m_\Lambda^3 (1+\beta) x (m_\Lambda^2 x^2 + Q^2)
\;\widetilde{\!\widetilde{B}}_6 (x) \nnb \\
\ar 8 e_s m_\Lambda^2 \Big\{ 2 m_s Q^2 \; \widehat{\!\widehat{B}}_6 -
m_\Lambda^2 m_s x^2 (\; \widehat{\!\widehat{C}}_6 - 2 \widehat{\!\widehat{B}}_8) 
- m_\Lambda (1+\beta) x \Big[ 2 (m_\Lambda^2 x^2 + Q^2)
\; \widehat{\!\widehat{B}}_6 \nnb \\
\ar m_\Lambda m_s x \; \widehat{\!\widehat{D}}_6 \Big]\Big\} (x)~, \nnb \\ \nnb \\
\rho_4^{\Lambda} (x)\es 
3 e_u m_\Lambda \Big\{
2 m_\Lambda^2 x \Big[ 2 (1-\beta) \; \check{\!\check{C}}_6
 - (1+\beta) (2 \; \check{\!\check{B}}_6 - 5 \;\check{\!\check{B}}_8) \Big] (x)
\nnb \\
\ek 3 \Big[ 2 (1-\beta) \Big( m_\Lambda^2 x^2
(\check{D}_5
- \check{C}_4 + 2 \check{C}_5 )
- Q^2 (\check{D}_2 - \check{C}_2) \Big) \nnb \\
\ar (1+\beta) \Big( Q^2 (3 \check{B}_2 + 7
\check{B}_4)  + m_\Lambda^2 x^2 ( 2 \check{H}_1
- 2 \check{E}_1 - \check{B}_2 +
\check{B}_4
- 10 \check{B}_5 - 20 \check{B}_7 ) \Big) \Big] (x) \nnb \\
\ar 2 m_\Lambda^2 x \int_0^{\bar{x}}dx_3\,\Big[ 2 (1-\beta) {V}_1^M + 5     
(1+\beta) T_1^M\Big] (x,1-x-x_3,x_3) \Big\} \nnb \\
\ar e_d m_\Lambda \Big\{
2 m_\Lambda^2 x \Big[ 2 (1-\beta) \; \widetilde{\!\widetilde{C}}_6
 - (1+\beta) (10 \; \widetilde{\!\widetilde{B}}_6 -
9 \; \widetilde{\!\widetilde{B}}_8) \Big] (x) \nnb \\
\ar \Big[ 2 (1-\beta) \Big( m_\Lambda^2 x^2
(\check{D}_5 + \check{C}_4 - 6 \check{C}_5 )
- Q^2 (\check{D}_2 + 5 \check{C}_2) \Big) \nnb \\
\ar (1+\beta) \Big( Q^2 (\check{B}_2 - 19
\check{B}_4)  + m_\Lambda^2 x^2 ( 2 \check{H}_1
- 2 \check{E}_1 + 5 \check{B}_2 -
5 \check{B}_4
+ 18 \check{B}_5 + 36 \check{B}_7 ) \Big) \Big] (x) \nnb \\
\ar  2 m_\Lambda^2 x \int_0^{\bar{x}} \,dx_1 \Big[ 2 (1-\beta) {V}_1^M + 9  
(1+\beta) T_1^M\Big] (x_1,x,1-x_1-x) \Big\} \nnb \\
\ar 2 e_s m_\Lambda \Big\{
4  m_\Lambda \Big[ m_\Lambda (1-\beta) x \;
\widehat{\!\widehat{C}}_6 -  m_\Lambda (1+\beta) x (2 \; \widehat{\!\widehat{B}}_6
- 3 \; \widehat{\!\widehat{B}}_8 ) + m_s \; \widehat{\!\widehat{B}}_6 \Big]
(x) \nnb \\ 
\ar \Big[ 2 (1-\beta) \Big(
m_\Lambda^2 x^2 (\widehat{D}_5 + \widehat{C}_4 - 3 \widehat{C}_5) -
Q^2 (\widehat{D}_2 + 2 \widehat{C}_2)\Big) \nnb \\
\ar 2 m_\Lambda m_s x ( 2 \widehat{C}_2 - \widehat{C}_4 -
\widehat{C}_5 - 4 \widehat{B}_4 + 4 \widehat{B}_5) \nnb \\
\ek (1+\beta) \Big( 2 Q^2 ( \widehat{B}_2 + 5  \widehat{B}_4) -
m_\Lambda^2 x^2 (\widehat{H}_1 - \widehat{E}_1 + \widehat{B}_2 - \widehat{B}_4 +
6 \widehat{B}_5 + 12 \widehat{B}_7) \nnb \\
\ek  m_\Lambda m_s x ( 2 \widehat{D}_2 + \widehat{D}_4 + \widehat{D}_5)
\Big) \Big] (x) \nnb \\
\ar 4 m_\Lambda^2 x \int_0^{\bar{x}} \,dx_1 \Big[ (1-\beta) {V}_1^M + 3  
(1+\beta) T_1^M\Big] (x_1,1-x_1-x,x) \Big\}\nnb \\ \nnb
\rho_2^{\Lambda}(x) \es
6 e_u m_\Lambda \Big\{
\Big[ (1-\beta) ( \check{D}_2 + \check{C}_2) -
(1+\beta) ( \check{B}_2 - \check{B}_4) \Big] (x) \nnb \\
\ar x \int_0^{\bar{x}}dx_3\,\Big[ (1-\beta) (A_3 + 2 {V}_1 - 3
{V}_3) - (1+\beta)(P_1 + S_1 - 5 T_1 + 10 T_3) \Big] (x,1-x-x_3,x_3) \Big\} \nnb \\
\ek 2 e_d m_\Lambda \Big\{
\Big[ (1-\beta) ( \widetilde{D}_2 + 3 \widetilde{C}_2) -
3 (1+\beta) ( \widetilde{B}_2 - \widetilde{B}_4) \Big] (x) \nnb \\
\ar x \int_0^{\bar{x}} \,dx_1 \Big[ (1-\beta) (A_3 - 2 {V}_1 + 7 {V}_3) -
(1+\beta)(P_1 + S_1 + 9 T_1 - 18 T_3) \Big] (x_1,x,1-x_1-x) \Big\}\nnb \\
\ek 4 e_s \Big\{
m_\Lambda (1-\beta) \widehat{D}_2 (x) \nnb \\
\ar \int_0^{\bar{x}}dx_1\,\Big[ m_\Lambda (1-\beta) x ( A_3 - 2 V_1 +
4{V}_3) - (1+\beta) \Big( m_\Lambda x (P_1+S_1+6 T_1-12T_3) + m_s A_1 \Big)\nnb \\
\ar 2 m_s ({V}_1 - 2 T_1)
\Big] (x_1,1-x_1-x,x) \Big\}~, \nnb \\ \nnb \\
\rho_6^{'\Lambda} (x) \es  12 e_u m_\Lambda^2 (1+\beta) (m_\Lambda^2 x^2 + Q^2)
\;\check{\!\check{B}}_6 (x) +
20 e_d m_\Lambda^2 (1+\beta) (m_\Lambda^2 x^2 + Q^2)
\;\widetilde{\!\widetilde{B}}_6 (x) \nnb \\
\ar 8 e_s m_\Lambda^2  \Big\{ m_\Lambda m_s x (\; \widehat{\!\widehat{C}}_6
+ \widehat{\!\widehat{B}}_6 - 2 \widehat{\!\widehat{B}}_8)
+ (1+\beta) \Big[ 2 (m_\Lambda^2 x^2 + Q^2) \; \widehat{\!\widehat{B}}_6 +
m_\Lambda m_s x \; \widehat{\!\widehat{D}}_6\Big]
\Big\} (x)~, \nnb \\ \nnb \\
\rho_4^{'\Lambda} (x) \es
3 e_u m_\Lambda^2 \Big\{
(1+\beta) \;\check{\!\check{B}}_6 (x) \nnb \\
\ar 2 x \Big[ (1-\beta) (\check{D}_2 + \check{D}_5 - \check{C}_2 -
\check{C}_4 + 2 \check{C}_5 ) \nnb \\
\ar
(1+\beta) (\check{H}_1 - \check{E}_1 - 2 \check{B}_2 - 3 \check{B}_4 - 5
\check{B}_5 - 10 \check{B}_7 ) \Big] (x) \nnb \\
\ar 2 (1-\beta) \int_0^{\bar{x}}dx_3\, (A_1^M - {V}_1^M)
(x,1-x-x_3,x_3)\Big\} \nnb \\
\ar e_d m_\Lambda^2 \Big\{
21 (1+\beta) \;\widetilde{\!\widetilde{B}}_6 (x) \nnb \\
\ek 2 x \Big[ (1-\beta) (\widetilde{D}_2 + \widetilde{D}_5 +
5 \widetilde{C}_2 + \widetilde{C}_4 - 6 \widetilde{C}_5 ) \nnb \\
\ar (1+\beta) (\widetilde{H}_1 - \widetilde{E}_1 + 2 \widetilde{B}_2 + 7 \widetilde{B}_4
+ 9 \widetilde{B}_5 + 18 \widetilde{B}_7 ) \Big] (x) \nnb \\
\ek 2  (1-\beta) \int_0^{\bar{x}}dx_1\, (A_1^M - 3 {V}_1^M)
(x_1,x,1-x_1-x) \Big\} \nnb \\
\ar 4  e_s  m_\Lambda \Big\{
3 m_\Lambda (1+\beta) \;\widehat{\!\widehat{B}}_6 (x) \nnb \\
\ek \Big[ (1-\beta) \Big( m_\Lambda x (\widehat{D}_2 +
\widehat{D}_5 + 2 \widehat{C}_2 + \widehat{C}_4 - 3 \widehat{C}_5)\Big) \nnb \\
\ar (1+\beta) \Big( m_\Lambda x (\widehat{H}_1 - \widehat{E}_1 +
2 \widehat{B}_2 + 4 \widehat{B}_4 + 6 \widehat{B}_5 + 12 \widehat{B}_7) +
m_s \widehat{D}_2\Big)
+ 2 m_s (\widehat{C}_2 - \widehat{B}_2 - \widehat{B}_4)
 \Big] (x) \nnb \\
\ek m_\Lambda (1-\beta) \int_0^{\bar{x}}dx_1\, {A}_1^M (x_1,1-x_1-x,x) \Big\}~,
\nnb \\ \nnb \\
\rho_2^{'\Lambda} (x) \es 6 e_u (1-\beta) \int_0^{\bar{x}}dx_3\, (A_1 - {V}_1)
(x,1-x-x_3,x_3) \nnb \\
\ek 2 e_d (1-\beta) \int_0^{\bar{x}}dx_1 \,(A_1 - 3 {V}_1) (x_1,x,1-x_1-x) \nnb \\
\ek 4 e_s (1-\beta) \int_0^{\bar{x}}dx_1\,{A}_1 (x,1-x-x_3,x_3)~,\nnb
\eea

In the above expressions for $\rho_i$ and $\rho_i^{'}$,
the functions ${\cal F}(x_i)$ are defined in the following way:

\bea
\label{eslff12}
\check{\cal F}(x_1) \es \int_1^{x_1}\!\!dx_1^{'}\int_0^{1- x^{'}_{1}}\!\!dx_3\,
{\cal F}(x_1^{'},1-x_1^{'}-x_3,x_3)~, \nnb \\
\check{\!\!\!\;\check{\cal F}}(x_1) \es 
\int_1^{x_1}\!\!dx_1^{'}\int_1^{x^{'}_{1}}\!\!dx_1^{''}
\int_0^{1- x^{''}_{1}}\!\!dx_3\,
{\cal F}(x_1^{''},1-x_1^{''}-x_3,x_3)~, \nnb \\
\widetilde{\cal F}(x_2) \es \int_1^{x_2}\!\!dx_2^{'}\int_0^{1- x^{'}_{2}}\!\!dx_1\,
{\cal F}(x_1,x_2^{'},1-x_1-x_2^{'})~, \nnb \\
\widetilde{\!\widetilde{\cal F}}(x_2) \es 
\int_1^{x_2}\!\!dx_2^{'}\int_1^{x^{'}_{2}}\!\!dx_2^{''}
\int_0^{1- x^{''}_{2}}\!\!dx_1\,
{\cal F}(x_1,x_2^{''},1-x_1-x_2^{''})~, \nnb \\
\widehat{\cal F}(x_3) \es \int_1^{x_3}\!\!dx_3^{'}\int_0^{1- x^{'}_{3}}\!\!dx_1\,
{\cal F}(x_1,1-x_1-x_3^{'},x_3^{'})~, \nnb \\
\widehat{\!\widehat{\cal F}}(x_3) \es 
\int_1^{x_3}\!\!dx_3^{'}\int_1^{x^{'}_{3}}\!\!dx_3^{''}
\int_0^{1- x^{''}_{3}}\!\!dx_1\,
{\cal F}(x_1,1-x_1-x_3^{''},x_3^{''})~,\nnb
\eea

and the definitions of the functions $B_i$, $C_i$, $D_i$, $E_1$ and $H_1$
are given as

\bea
\label{eslff13}
B_2 \es T_1+T_2-2 T_3~, \nnb \\
B_4 \es T_1-T_2-2 T_7~, \nnb \\
B_5 \es - T_1+T_5+2 T_8~, \nnb \\
B_6 \es 2 T_1-2 T_3-2 T_4+2 T_5+2 T_7+2 T_8~, \nnb \\
B_7 \es T_7-T_8~, \nnb \\
B_8 \es  -T_1+T_2+T_5-T_6+2 T_7+2T_8~, \nnb \\
C_2 \es V_1-V_2-V_3~, \nnb \\
C_4 \es -2V_1+V_3+V_4+2V_5~, \nnb \\
C_5 \es V_4-V_3~, \nnb \\
C_6 \es -V_1+V_2+V_3+V_4+V_5-V_6~, \nnb \\
D_2 \es -A_1+A_2-A_3~, \nnb \\
D_4 \es -2A_1-A_3-A_4+2A_5~, \nnb \\
D_5 \es A_3-A_4~, \nnb \\
D_6 \es A_1-A_2+A_3+A_4-A_5+A_6~, \nnb \\
E_1 \es S_1-S_2~, \nnb \\
H_1 \es P_2-P_1~. \nnb
\eea

The expressions of the functions $\rho_i(x)$ and $\rho_i^{'}(x)$ for the
$\Xi^0~(\Xi^-)$ baryons can be obtained from the corresponding results of
$\Sigma^+~(\Sigma^-)$ by making the replacement $s \lrar u~(s \lrar d)$.

\newpage

\section*{Figure captions}
{\bf Fig. (1)} The dependence of the magnetic form factor $G_M(Q^2)$ of the
$\Sigma^+$ baryon on $Q^2$ at $s_0=3.0~GeV^2$ and $M^2=2.0~GeV^2$, at
several different fixed values of the arbitrary parameter $\beta$.\\ \\
{\bf Fig. (2)} The same as in Fig. (1), but for the electric charge
form factor $G_E(Q^2)$.\\ \\
{\bf Fig. (3)} The same as in Fig. (1), but for the $\Xi^-$ baryon at
$s_0=3.2~GeV^2$.\\ \\
{\bf Fig. (4)} The same as in Fig. (2), but for the $\Xi^-$ baryon at
$s_0=3.2~GeV^2$.\\ \\
{\bf Fig. (5)} The same as in Fig. (1), but for the $\Lambda$ baryon at
$s_0=2.6~GeV^2$ and $M^2=1.8~GeV^2$.\\ \\
{\bf Fig. (6)} The same as in Fig. (5), but for the electric charge
form factor $G_E(Q^2)$.\\ \\
{\bf Fig. (7)} The dependence of the magnetic form factor $G_M$ of the
$\Sigma^+$ baryon on $\cos\theta$ at $Q^2=1.0~GeV^2$ and $s_0=3.0~GeV^2$,
at several different fixed values of the Borel mass parameter $M^2$.\\ \\
{\bf Fig. (8)} The same as in Fig. (7), but for the electric charge
form factor $G_E$.

\newpage

\begin{figure}
\vskip 3. cm
    \includegraphics{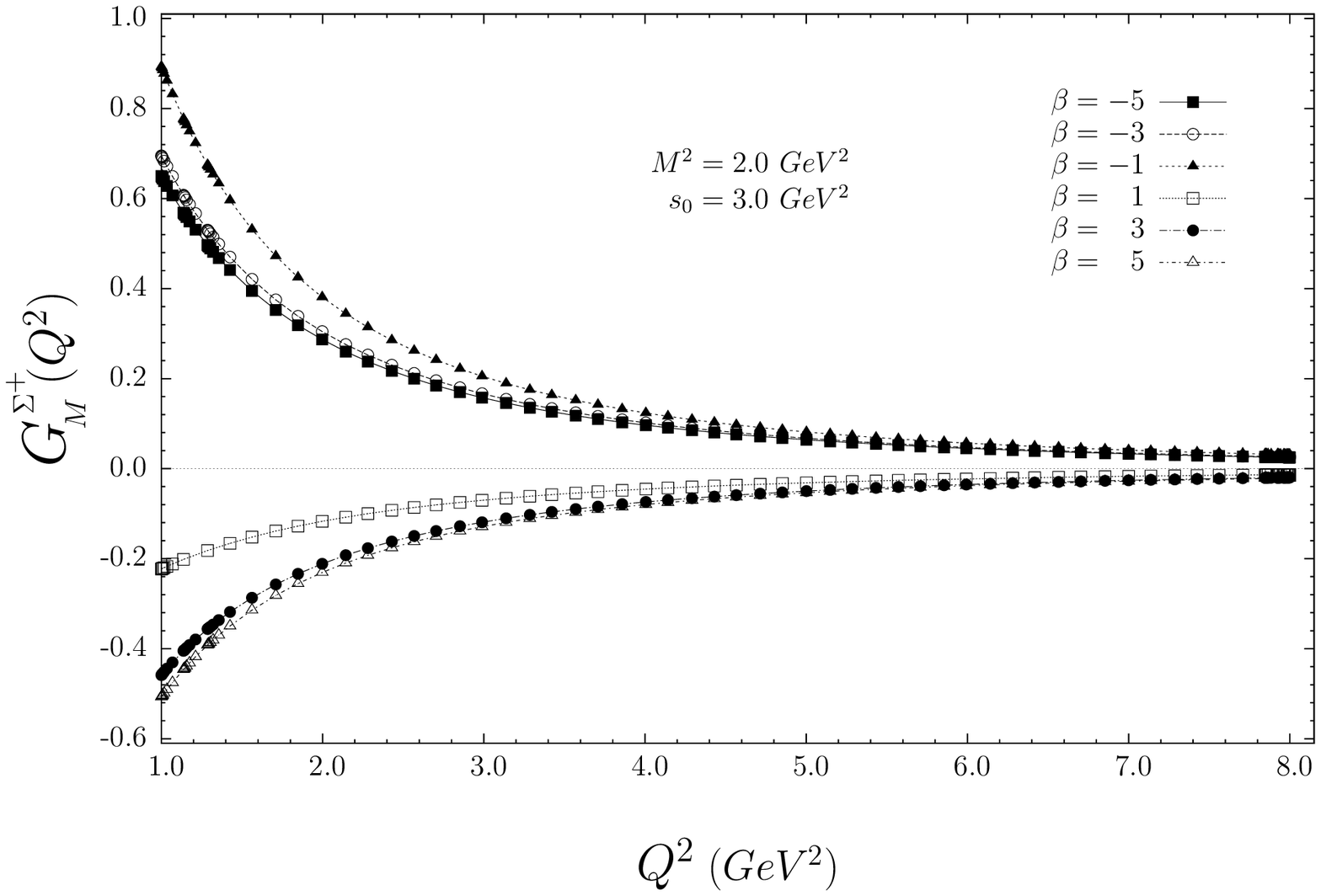}
\vskip 7.0cm
\caption{}
\end{figure}

\begin{figure}
\vskip 3. cm
    \includegraphics{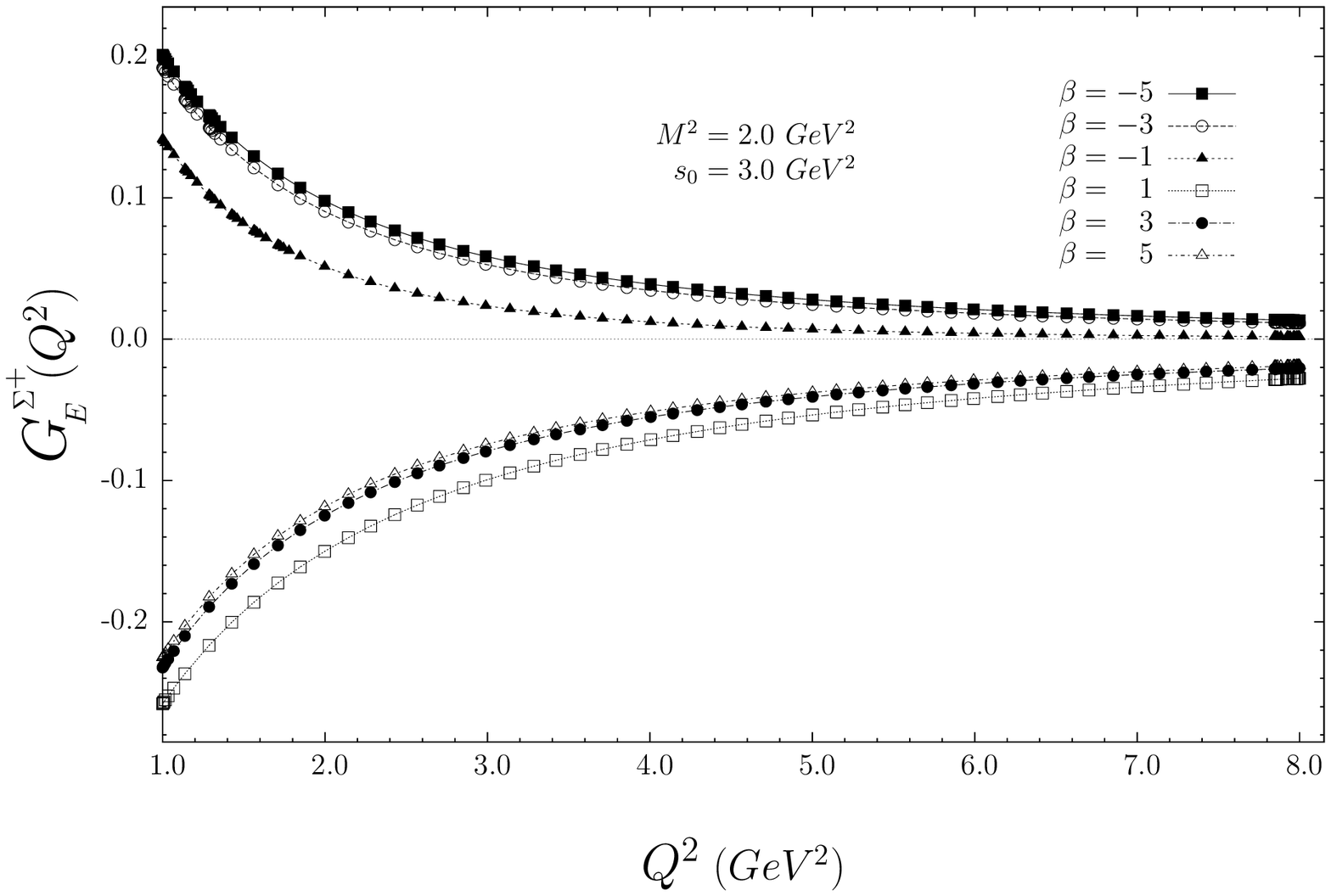}
\vskip 7.0cm
\caption{}
\end{figure}

\begin{figure}
\vskip 3. cm
    \includegraphics{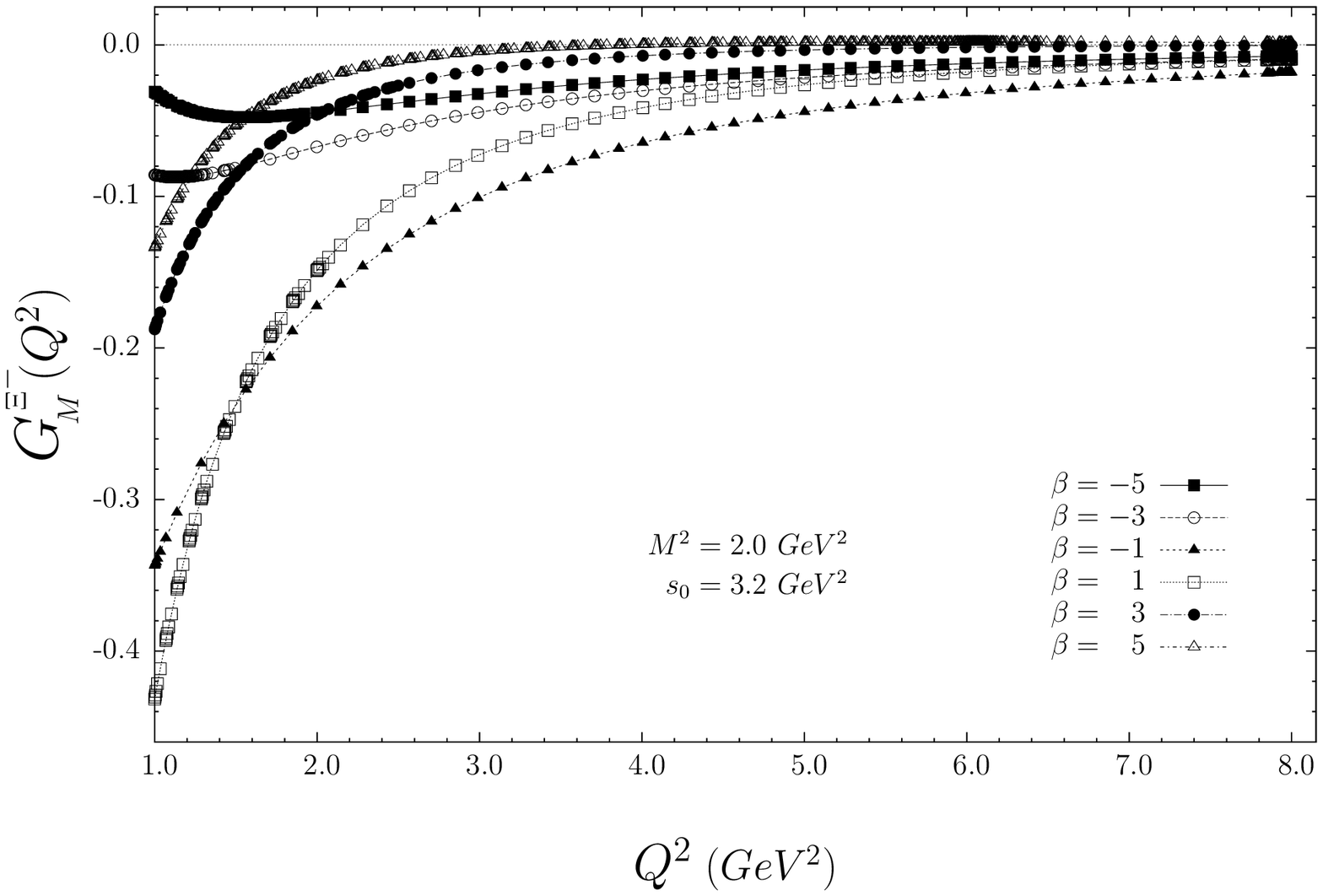}
\vskip 7.0cm
\caption{}
\end{figure}

\begin{figure}
\vskip 3. cm
    \includegraphics{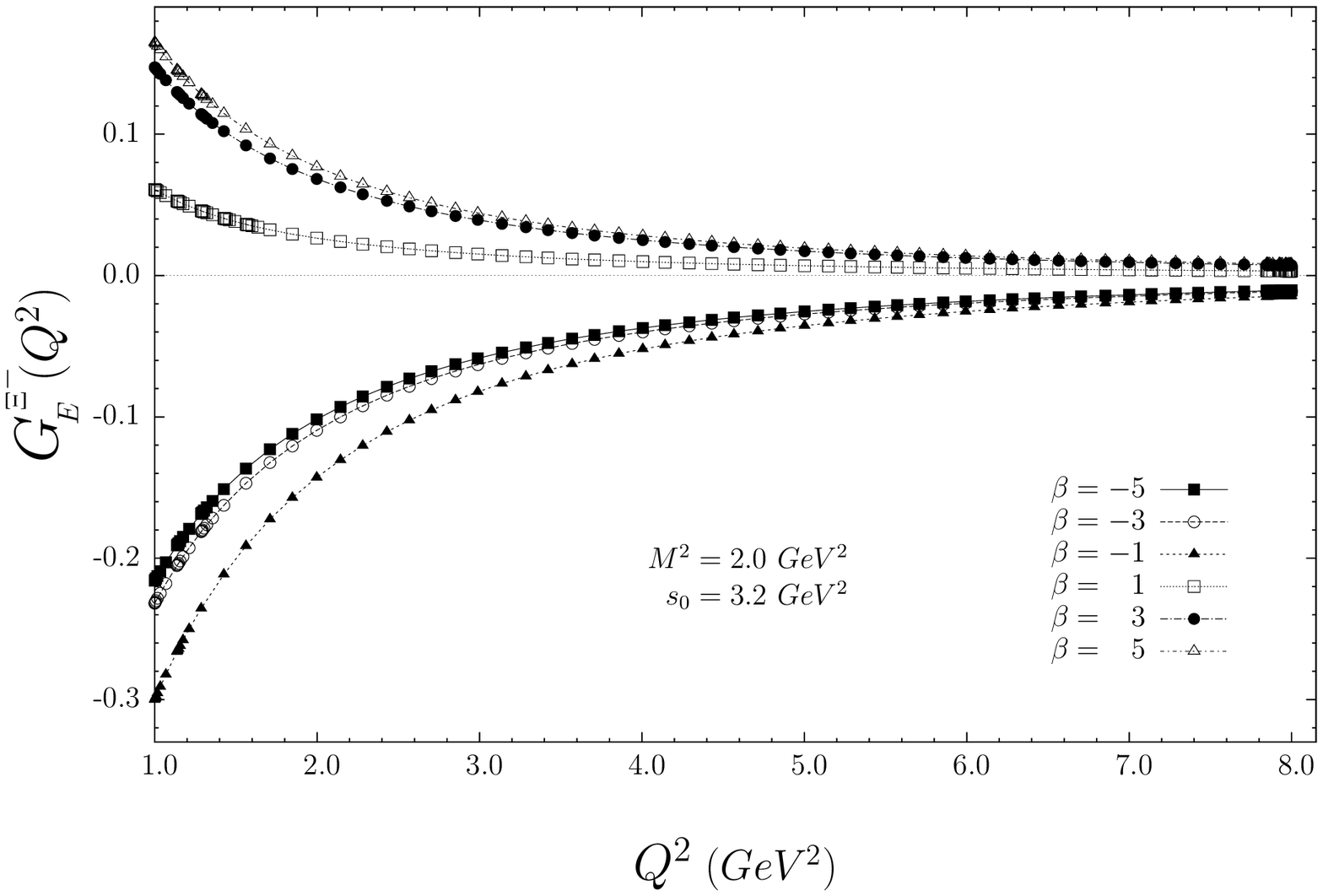}
\vskip 7.0cm
\caption{}
\end{figure}

\begin{figure}
\vskip 3. cm
    \includegraphics{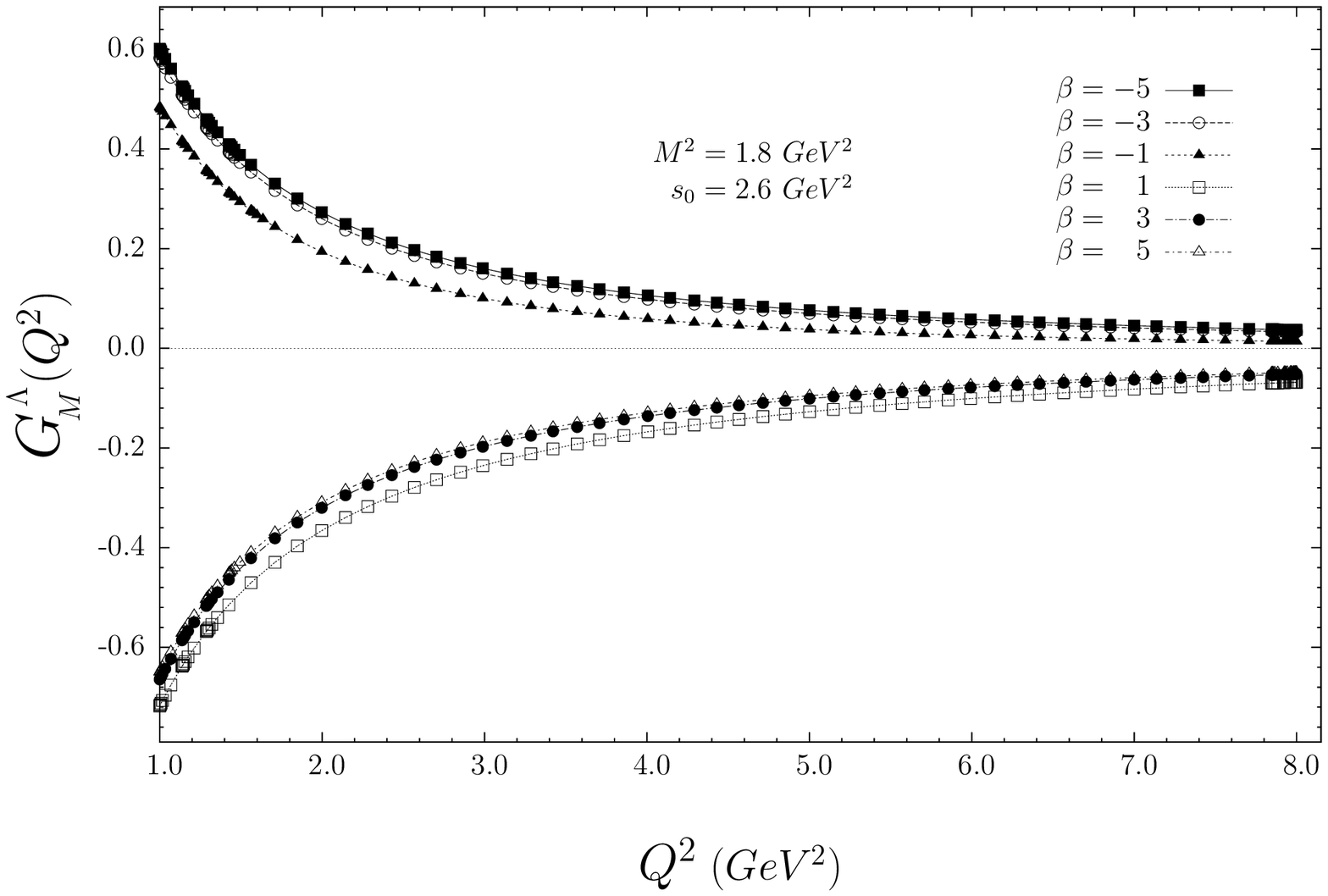}
\vskip 7.0cm
\caption{}
\end{figure}

\begin{figure}
\vskip 3. cm
    \includegraphics{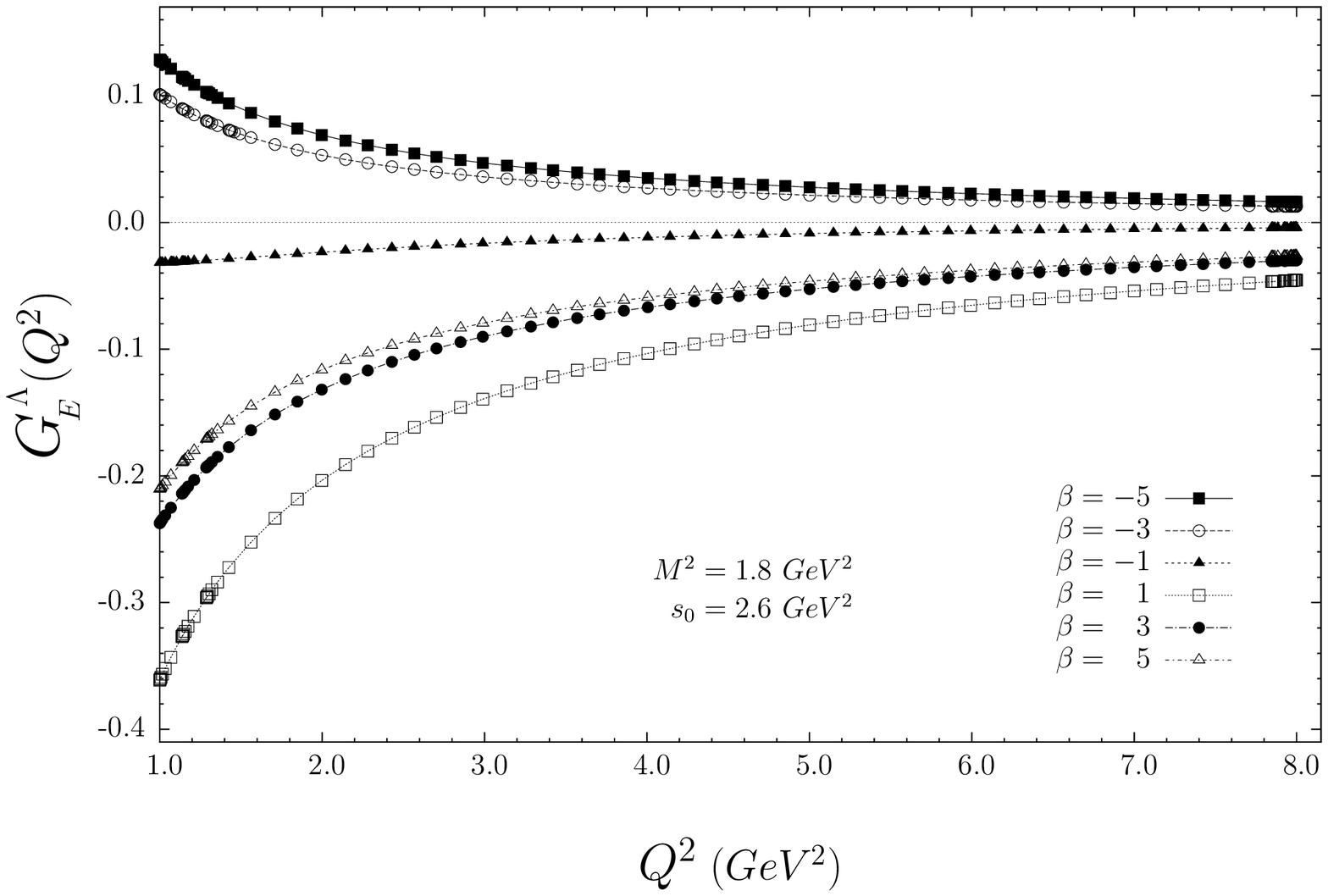}
\vskip 7.0cm
\caption{}
\end{figure}

\begin{figure}
\vskip 3. cm
    \includegraphics{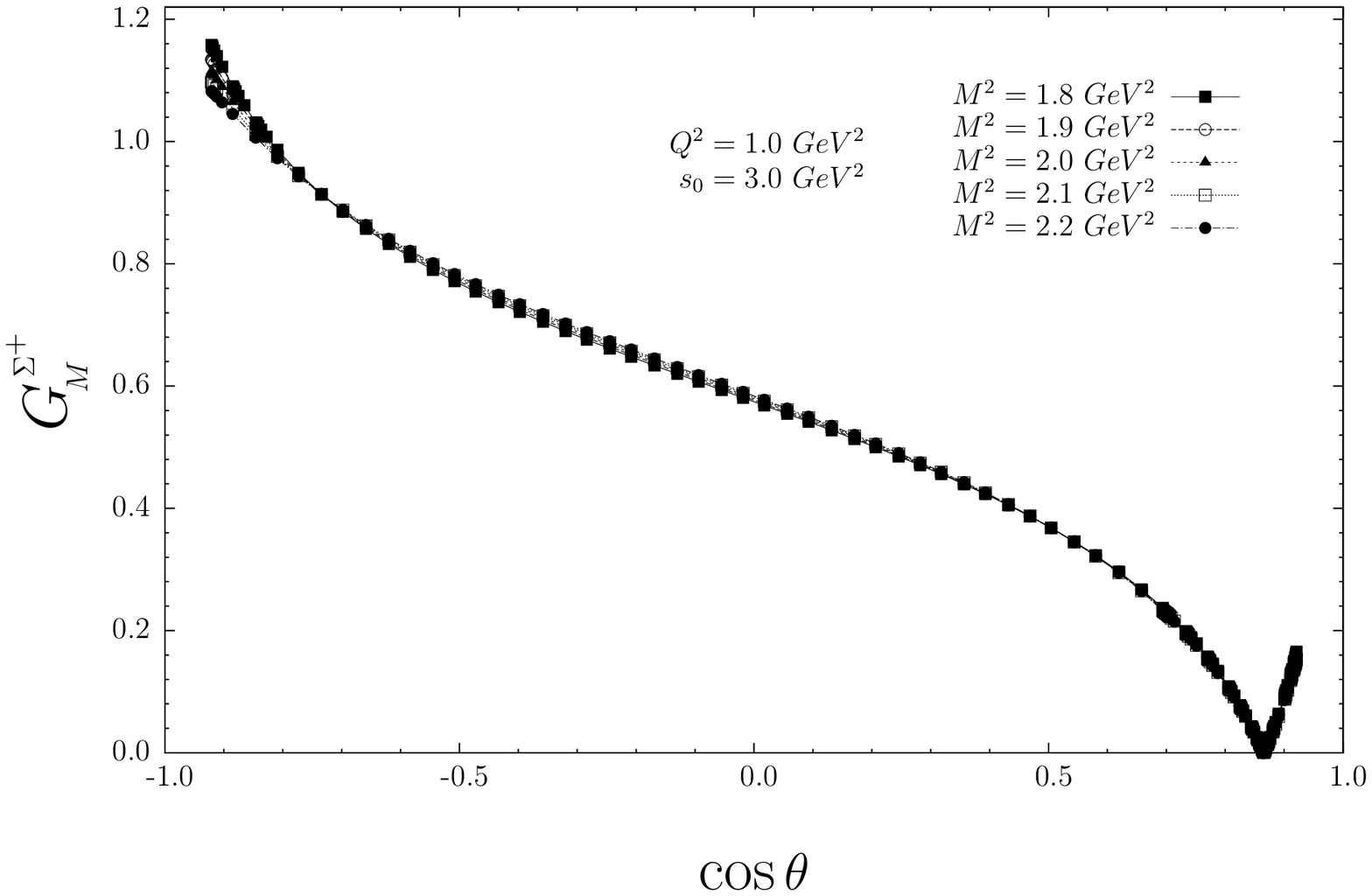}
\vskip 7.0cm
\caption{}
\end{figure}

\begin{figure}
\vskip 3. cm
    \includegraphics{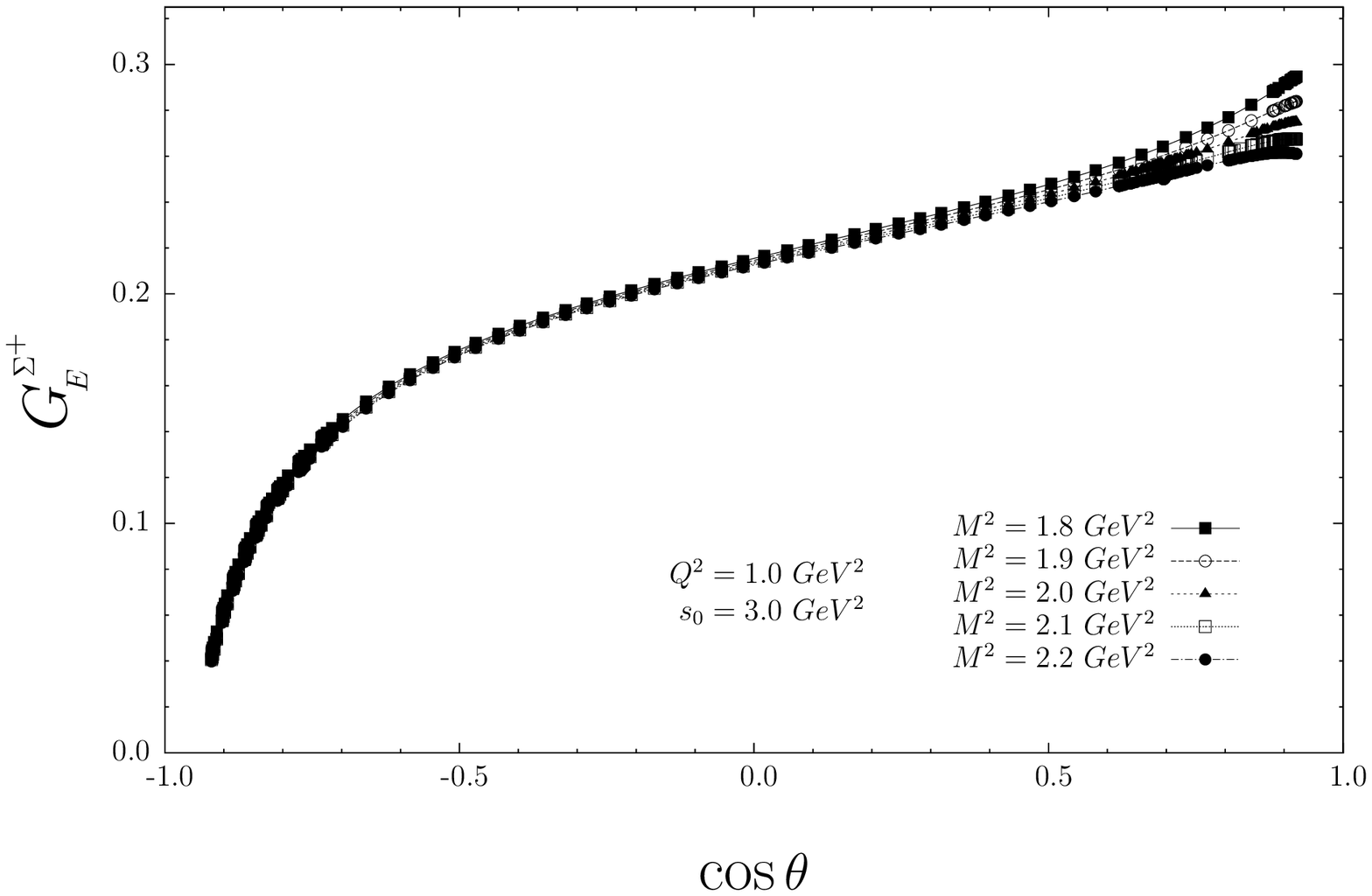}
\vskip 7.0cm
\caption{}
\end{figure}


\begin{thebibliography}{99}

\bibitem{Rgegm01} J. Arrington, C. D. Roberts, and J. M. Zanotti,
  J. Phys. G {\bf 34}, 523 (2007).

\bibitem{Rgegm02} A. J. R. Puckett {\it et. al},
  Phys. Rev. Lett. {\bf 104}, 242301 (2010);
  Phys. Rev. C {\bf 85}, 045203 (2012). 

\bibitem{Rgegm03} Q. Gayou {\it et. al},
  Phys. Rev. C {\bf 64}, 038202 (2001).

\bibitem{Rgegm04} V. Punjabi {\it et. al},
  Phys. Rev. C {\bf 71}, 055202 (2005).

\bibitem{Rgegm05} J. Lachniet {\it et. al}, CLAS Collaboration,
  Phys. Rev. Lett. {\bf 102}, 192001 (2009).

\bibitem{Rgegm06} S. Riordan {\it et. al},
  Phys. Rev. Lett. {\bf 105}, 262302 (2010).

\bibitem{Rgegm07} L. Tiator, D. Dreschel, S. S. Kamalov, and M. Vanderhaeghen,
  Eur. Phys. J. ST {\bf 198}, 141 (2011).

\bibitem{Rgegm08} Yang--Lu Liu, Ming--Qiu Huang,
  J. Phys. G {\bf 37}, 115010 (2010).

\bibitem{Rgegm09} Yang--Lu Liu, Ming--Qiu Huang,
  Phys. Rev. D {\bf 80}, 055015 (2009).

\bibitem{Rgegm10} Yang--Lu Liu, Ming--Qiu Huang,
  Phys. Rev. D {\bf 79}, 114031 (2009).

\bibitem{Rgegm11} V. M. Braun, A. Lenz and M. Wittmann,
  Phys. Rev. D {\bf 73}, 094019 (2006);
                  V. M. Braun, A. Lenz, N. Mahnke, and E. Stern, 
  Phys. Rev. D {\bf 65}, 074011 (2002).

\bibitem{Rgegm12} T. M. Aliev, K. Azizi, A. \"{O}zpineci, and M. Savc{\i},
  Phys. Rev. D {\bf 77}, 114014 (2008).

\bibitem{Rgegm13} A. \"{O}zpineci, S. Yakovlev, V. Zamiralov,
  Mod. Phys. Lett. A {\bf 20}, 243 (2005).

\bibitem{bir} V. M. Braun, R. J. Fries, N. Mahnke and E. Stein, Nucl. Phys. B {\bf 589}, 381 (2000).
\bibitem{iki} A. Lenz, M. G\"{o}ckeler, T. Kaltenbrunner and N. Warkentin, Phys. Rev D {\bf 79}, 093007 (2009).

\bibitem{Rgegm14} T. M. Aliev, A. \"{O}zpineci, and M. Savc{\i},
  Phys. Rev. D {\bf 66}, 016002 (2002);
  Erratum--ibid, D {\bf 67}, 039901 (2003).

\bibitem{Rgegm15} T. Van Cauteren, D. Merten, T. Corthals, S. Janssen,
                  B. Metsch, H. R. Petry, and J. Ryckebusch,
  Eur. Phys. J. A {\bf 20}, 283 (2004).

\bibitem{Rgegm16} G. Ramalho and  K. Tsushima,
  Phys. Rev. D {\bf 84}, 051301 (2011); G. Ramalho,  K. Tsushima and A. W. Thomas, J. Phys. G: Nucl. Part. Phys. {\bf 40},  015102 (2013). 



\bibitem{Rgegm17} Huey--Wen Lin, K. Orignos,
  Phys. Rev. D {\bf 79}, 074507 (2009).


\end{thebibliography}
\end{document}